\documentclass{svproc}
\usepackage{url}

\usepackage{float}
\usepackage[dvipdfmx]{graphicx}        
\usepackage{latexsym}

\begin{document}
\mainmatter              
\title{MEGADOCK-GUI: a GUI-based complete cross-docking tool for exploring protein-protein interactions}
\titlerunning{MEGADOCK-GUI}  
%
\author{Masahito Ohue\inst{1,3} \and Yutaka Akiyama\inst{2,3}}
\authorrunning{Masahito Ohue et al.} 
%
%
\institute{Department of Computer Science, School of Computing, Tokyo Institute of Technology, G3-56, 4259 Nagatsutacho, Midori-ku, Yokohama City, Kanagawa 226-8503, Japan\\ \email{ohue@c.titech.ac.jp}
\and
Department of Computer Science, School of Computing, Tokyo Institute of Technology, W8-76, 2-12-1 Ookayama, Meguro-ku, Tokyo 152-8550, Japan\\ \email{akiyama@c.titech.ac.jp}
\and
Ahead Biocomputing, Co. Ltd., Kawasaki Frontier Bldg. 4F, 11-2 Ekimaehoncho, Kawasaki-ku, Kawasaki City, Kanagawa 210-0007, Japan\\
\email{\{ohue, akiyama\}@ahead-biocomputing.co.jp}
}

\maketitle              

\begin{abstract}
Information on protein-protein interactions (PPIs) not only advances our understanding of molecular biology but also provides important clues for target selection in drug discovery and the design of PPI inhibitors. One of the techniques used for computational prediction of PPIs is protein-protein docking calculations, and a variety of software has been developed.
However, a friendly interface for users who are not sufficiently familiar with the command line interface has not been developed so far.
In this study, we have developed a graphical user interface, MEGADOCK-GUI, which enables users to easily predict PPIs and protein complex structures. In addition to the original 3-D molecular viewer and input file preparation functions, MEGADOCK-GUI is software that can automatically perform complete cross-docking of $M$ vs. $N$ proteins.
With MEGADOCK-GUI, various applications related to the prediction of PPIs, such as ensemble docking that handles multiple conformations of proteins and screening of binding partner proteins that bind to specific proteins, can now be easily performed.

\keywords{protein-protein interaction (PPI), protein-protein docking, MEGADOCK, MEGADOCK-GUI}
\end{abstract}
\section{Introduction}
Information on protein-protein interactions (PPIs) provides useful insights from biological understanding to drug target selection and PPI inhibitor design~\cite{PPI1,PPI2}.
Due to the enormous cost of experimentally determining PPIs, computational prediction techniques are becoming more important~\cite{PPIPred}.

Protein-protein docking (PPD) is a computational technique that predicts the structure of complexes of multiple proteins based on their tertiary structures~\cite{PPD1,PPD2}.
Just as protein-ligand docking techniques are used to screen for small-molecule compounds that bind to target proteins as well as predict binding poses, techniques exist to predict protein pairs that will interact by PPD~\cite{Matsuzaki,Wass,Zhang,MEGADOCK}.
In the past, even a PPD of single protein pair required a large amount of computational time, but in recent years, there have been advances in fast PPD methods and computational acceleration techniques~\cite{Accel1,Accel2,Accel3}.
The PPI network can now be predicted exhaustively by complete cross-docking with high-performance computing~\cite{O1,O2,Hex}.

However, these tools are technologies that reach only a limited number of users who can use parallel computers and Linux environments without any inconvenience. In this study, we developed MEGADOCK-GUI, a graphical user interface (GUI) environment for interactive and intuitive execution of MEGADOCK, one of the PPD software. MEGADOCK-GUI is written in Java and runs on multiple platforms. MEGADOCK-GUI can automatically perform 1 vs. 1 docking calculations as well as $M$ vs. $N$ complete cross-docking (all-against-all docking).

\section{Related Work}
As GUI tools for PPD, Hex~\cite{Hex} and InteractiveROSETTA (RosettaDock)~\cite{IRosetta} already exist.
In addition, some software sold as commercial tools provide an integrated GUI environment and PPD functions (e.g., CCG MOE~\cite{MOE}, Accelrys Discovery Studio~\cite{DS}, Schr\"{o}dinger PIPER~\cite{Schro}, and Molsoft ICM-Pro~\cite{ICM}).
However, except for Hex, there is no automatic complete cross-docking function. Also, for Hex, complete cross-docking is not fully automatic, and the last update was in 2013, so it is not continuously supported. Currently, MEGADOCK-GUI is the only software for complete cross-docking.

\section{MEGADOCK Overview}
MEGADOCK~\cite{MEGADOCK,MEGADOCK4} is our software for PPD and PPI prediction. The 3-D structures (Protein Data Bank (PDB) data) of two proteins (receptor protein and ligand protein) for predicting interaction are input, and the presence or absence of the interaction is output in the form of a score. The main part of the calculation is grid-based docking of the protein, which is implemented using FFT~\cite{Katchalski1992}. The FFT calculation depends on the protein size but is approximately 80\%{} of the total occupancy. The computational scale is $\mathcal{O}(N_v^3 \log N_v)$ if the size of one side of the grid is $N_v$, usually representing a protein in a grid of 1.2 \AA{} pitches.

MEGADOCK is a multi-threaded implementation that uses OpenMP and runs on a multi-core CPU. Furthermore, a GPU-implemented version is available, which runs on the multiple GPUs using the CUDA library~\cite{Accel1}. A multi-node parallel implementation version was also created by hybrid parallelization combined with MPI parallelization~\cite{Accel2,MEGADOCK4}.

\section{Software Implementation and Features}
\subsection{Implementation}
MEGADOCK is written in C++ and runs in a normal Linux environment. MEGADOCK-GUI was constructed as a wrapper tool for executing MEGADOCK binary files. The outline of MEGADOCK-GUI is shown in Fig.~1. MEGADOCK-GUI is a GUI environment for preparing input files and setting parameters necessary for the operation of MEGADOCK, and includes an original 3-D molecular viewer for visual analysis of PDB files and MEGADOCK result files. MEGADOCK version 4.0~\cite{MEGADOCK4} was used in this study.

The user interface and functions of the MEGADOCK-GUI were implemented using the Java language and the Eclipse IDE~\cite{eclipse}. 64 bit Linux or 64/32 bit Windows 10 is required to run the MEGADOCK-GUI, and Java Runtime Environment Version 7 or later must be installed.


\begin{figure}[tb]
  \begin{center}
    \includegraphics[width=.8\textwidth]{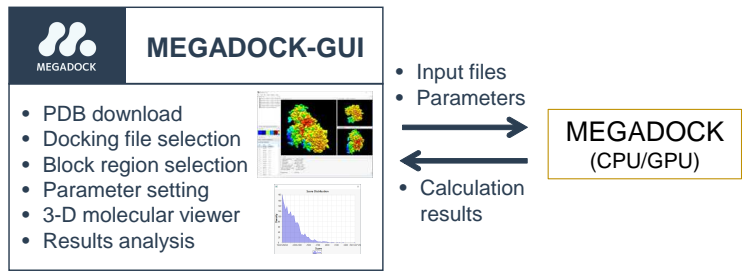}
    \caption{Overview of MEGADOCK-GUI and its relationship with MEGADOCK}
    \label{Fig1}
  \end{center}
\end{figure}

\begin{figure}[tb]
  \begin{center}
    \includegraphics[width=\textwidth]{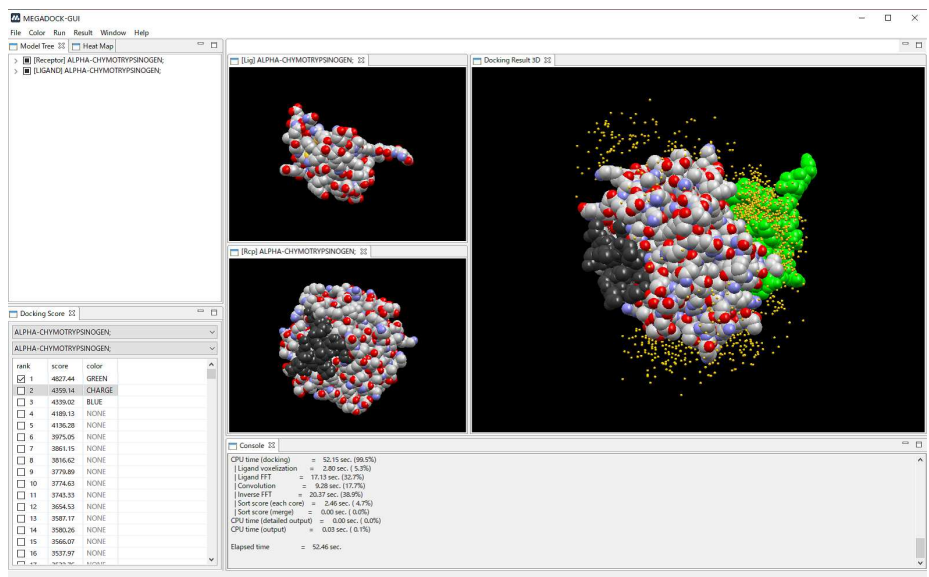}
    \caption{The view of MEGADOCK-GUI. Receptor, ligand, and docking results are displayed in sub-windows.Each sub-window is a molecular viewer, which allows the user to move and rotate the protein using the mouse.  The black areas on the protein are the residues designated as undocked regions. The centroids (yellow) of the docking results are almost non-existent on undocked region.}
    \label{Fig2}
  \end{center}
\end{figure}

\subsection{Features}
MEGADOCK-GUI implements functions to support the execution of docking calculations and analysis of the results in MEGADOCK. The main functions of MEGADOCK-GUI are listed below.

\subsubsection{PDB file download function}
In addition to reading PDB files from the local storage, it also has a function to download PDB files from PDBj~\cite{pdbj}, which are the input for docking calculations.
\subsubsection{Chain splitting function}
When there are multiple chains in a PDB file, it is possible to control whether or not each chain is included in the PPD input.
\subsubsection{Setting of undocking region}
There is a function to set an undocking region (a constraint that prevents binding around the residue) for each amino acid residue. The user can set the undocking region by checking or unchecking the checkbox on the list of amino acid residues or by double-clicking the mouse directly on the viewer (Fig.~2).
\subsubsection{Setting docking calculation parameters}
In the dialog box, you can change the parameters required for the MEGADOCK docking calculation. The main parameters include the number of candidate structures to be output, the tick angle for rotation sampling, and the weights for the strength of the electrostatic interaction term and hydrophobic term.
\subsubsection{Automatic execution of $M$ vs. $N$ docking calculations}.
By preparing multiple receptors and ligands as input files, $M$ vs. $N$ docking calculations can be performed automatically. The docking scores of the results of the $M$ vs. $N$ docking calculation can be listed as an $M\times N$ matrix-like heat map.
\subsubsection{Setting up the computing environment}
As well as the docking calculation parameters, the calculation environment can also be changed via a dialog box. For example, the user can change the number of CPU cores to be used, or change the MEGADOCK binary file, etc.
\subsubsection{3-D molecular viewer}
Each input file and output result can be displayed in the 3-D molecular viewer.
The viewer can be launched as multiple sub-windows, and the position, size, and the number of sub-windows can be changed (Fig.~2). As for the docking results, multiple candidate structures can be displayed in different colors, and plots of the centroid of ligand proteins can be displayed to get an overview of the interaction sites.
\subsubsection{Docking score analysis}
For each docking pair, the density distribution of docking scores by MEGADOCK can be displayed (Fig.~3).









\begin{figure}[tb]
  \begin{center}
    \includegraphics[width=.5\textwidth]{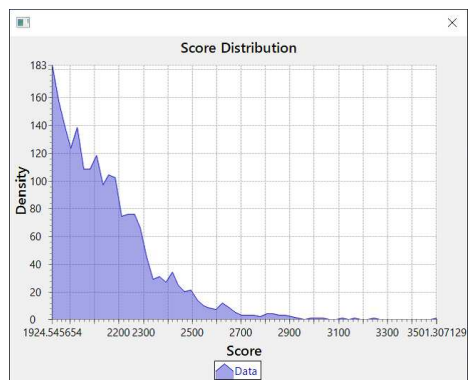}
    \caption{Density distribution of docking scores}
    \label{Fig3}
  \end{center}
\end{figure}

\section{Examples of Software Use}
This section describes an example of using the MEGADOCK-GUI.
The PDB files required as input for MEGADOCK can be obtained by selecting them from local storage or by downloading them by entering the PDB ID (Fig.~4).
In Fig.~5, PDB ID: 2DCY (Xylanase) and PDB ID: 1T6E (Xylanase inhibitor) are specified as Receptor and Ligand, respectively.
Since 2DCY has five chains registered as asymmetric unit~\cite{Xylanase1}, five protein chains are displayed as Receptors (Fig.~5).

\begin{figure}[tb]
  \begin{center}
    \includegraphics[width=.4\textwidth]{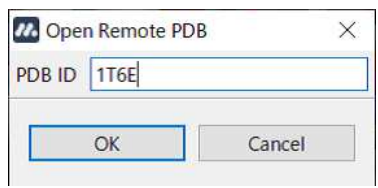}
    \caption{PDB file download function using MEGADOCK-GUI, which allows the user to download the protein structure of interest by entering the PDB ID.}
    \label{Fig4}
  \end{center}
\end{figure}

\begin{figure}[tb]
  \begin{center}
    \includegraphics[width=\textwidth]{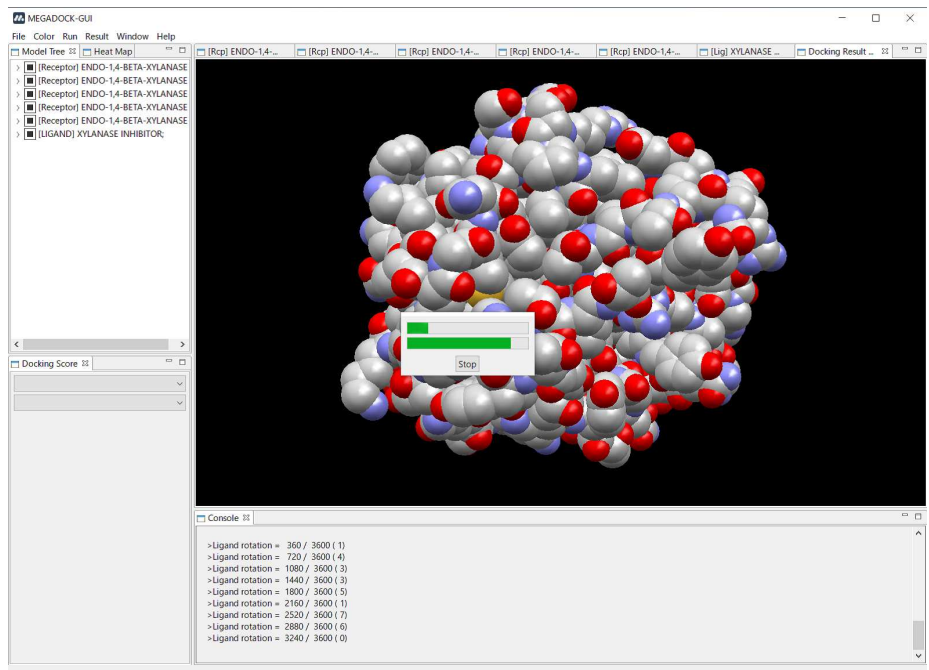}
    \caption{MEGADOCK-GUI performing $5:1$ docking calculation with PDB ID: 2DCY (Xylanase) as Receptor and PDB ID: 1T6E (Xylanase inhibitor) as Ligand.}
    \label{Fig5}
  \end{center}
\end{figure}

When MEGADOCK is executed from the `Run' menu in this state, 5 to 1, or 5 docking calculations will be executed automatically.
The two progress bars allow the user to grasp the current progress of the docking calculation.

When the docking calculation is completed, the results of the docking calculation can be displayed.
In MEGADOCK-GUI, the user can plot the ligand centroids by the points shown in Fig.~2, and display each candidate structure one by one.
In addition to the single color and CPK coloring (Fig.~5), the proteins can also be displayed by their charge distribution, as shown in Fig.~6. In Fig.~6, we can see that the complex structure is obtained by binding to the characteristic surface charge of Ligand protein.
The heat map in the center-left of the screen in Fig.~6 shows the results of the interaction prediction by MEGADOCK, and the darker the red color, the stronger the possibility of PPI.
Although Xylanase and Xylanase inhibitor are already known to interact with each other~\cite{Xylanase1,Xylanase2}, the prediction of the interaction differs depending on the conformation of the chain structure. This suggests the usefulness of ensemble docking in which multiple conformations are used together.

\begin{figure}[H]
  \begin{center}
    \includegraphics[width=\textwidth]{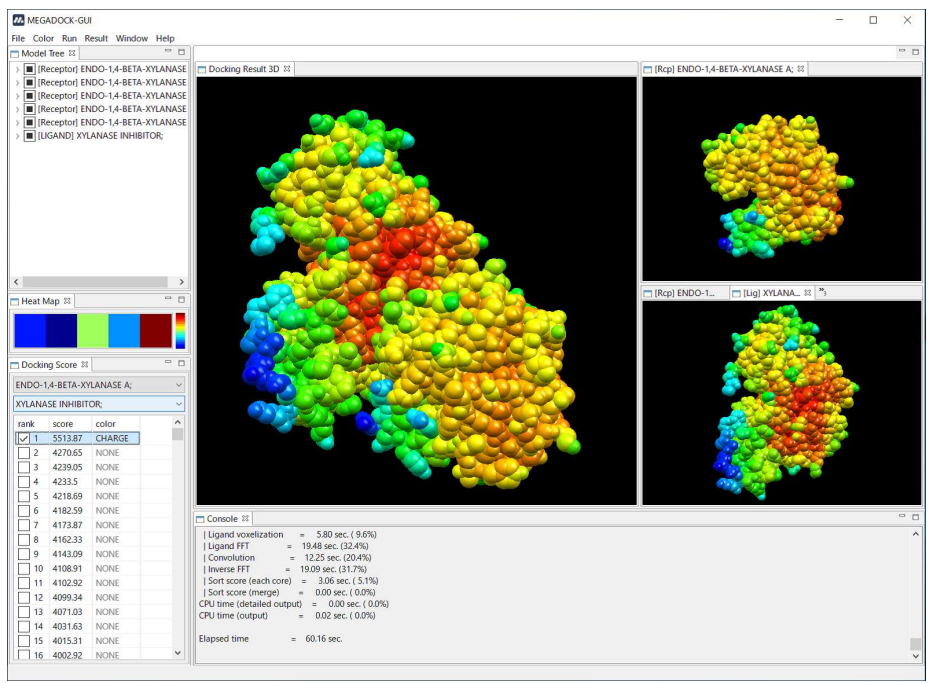}
    \caption{Screenshot of the results of MEGADOCK calculations of the five chains of Xylanase and the Xylanase inhibitor. Each protein is color-coded according to its surface charge value. The heat map in the center left shows the visualization of the PPI score values obtained from the five docking calculations.}
    \label{Fig6}
  \end{center}
\end{figure}

\section{Conclusions}
In this study, we have developed MEGADOCK-GUI, which can perform PPD calculations via a user-friendly GUI.
MEGADOCK-GUI has an interface especially for rapid docking calculations for multiple pairs, and can automatically perform complete cross-docking of $M$ vs. $N$.
With MEGADOCK-GUI, various applications related to the prediction of PPIs, such as ensemble docking for multiple conformations of a protein, and screening of binding partner proteins that bind to a specific protein, can be easily performed. 

MEGADOCK-GUI assumes the use of a local computing environment, but if it becomes possible to deploy calculations using a public cloud environment such as the Amazon Elastic Computing Cloud (EC2), the scale of calculations can be expanded and the range of applications widened.
MEGADOCK is already capable of parallelization in a public cloud environment~\cite{Cloud1,Cloud2,Cloud3}, and implementing an automatic execution function in MEGADOCK-GUI is a future task. In addition, additional implementation of the post-docking analysis function~\cite{Frontier,Uchikoga} should be considered to support users' research.


\section*{Acknowledgements}
This work was partially supported by the Japan Society for the Promotion of Science (JSPS) KAKENHI (20H04280), and the Platform Project for Supporting Drug Discovery and Life Science Research (Basis for Supporting Innovative Drug Discovery and Life Science Research (BINDS)) (Grant No. JP20am0101112) from the Japan Agency for Medical Research and Development (AMED).

%
%

\end{document}